\documentclass[twocolumn,showpacs,preprintnumbers,amsmath,amssymb,fleq]{revtex4}

\usepackage{graphicx}
\usepackage{dcolumn}
\usepackage{bm}

\begin{document}

\preprint{Published in ArXiv/Physics/Optics}

\title{Theory of optical imaging beyond the diffraction limit
with a far-field superlens}

\author{St\'ephane Durant}
\email{stephane.durant@gmail.com}
\author{Zhaowei Liu}%
\author{Nicholas Fang}
 \altaffiliation[Present address ]{Department of Mechanical and Industrial
 Engineering, University of Illinois at Urbana-Champaign 158 Mechanical
Engineering Building, MC-244 1206 West Green Street
Urbana, IL 61801}
\author{Xiang Zhang}
\email{xiang@berkeley.edu}

\affiliation{%
5130 Etcheverry Hall, NSF Nanoscale Science and Engineering Center
(NSEC)\\  University of California, Berkeley CA 94720-1740, USA}%


\date{May 23rd 2005, last revision \today }

\begin{abstract}
Recent theoretical and experimental studies have shown that
imaging with resolution well beyond the diffraction limit can be
obtained with so-called superlenses. Images formed by such
superlenses are, however, in the near field only, or a fraction of
wavelength away from the lens. In this paper, we propose a
far-field superlens (FSL) device which is composed of a planar
superlens with periodical corrugation. We show in theory that when
an object is placed in close proximity of such a FSL, a unique
image can be formed in far-field. As an example, we demonstrate
numerically that images of 40 nm lines with a 30 nm gap can be
obtained from far-field data with properly designed FSL working at
376nm wavelength.
\end{abstract}


\maketitle

\section{Introduction}
Conventional optical lens imaging suffers from the diffraction
limit which originates from the loss of evanescent waves that
cannot travel to the far field. In the near-field, an object under
illumination scatters both propagating and evanescent waves.
Propagating waves carry the low spatial resolution information on
the field modulation up to $\lambda_{0}/n$, where $\lambda_{0}$ is
the illumination wavelength in vacuum and $n$ is the refractive
index of the surrounding medium. On the other hand, information on
modulation of the field smaller than $\lambda_{0}/n$ are carried
by evanescent waves.

One promising approach of imaging beyond the diffraction limit
emerged with the recent proposal of superlenses\cite{Pendry00}.
Basically a superlens is made of a slab of material that can
support surface waves along the slab from electromagnetic
excitation. In contrast with conventional material for which
evanescent waves decays, using a superlens, evanescent waves can
be transmitted with enhanced amplitude resulting from the surface
wave excitation. Therefore a superlens has the ability to
effectively recover the evanescent components radiated by the
object in the image plane\cite{Smith05}\cite{Fang05}. A good image
can be obtained if the enhancement of evanescent waves by the
superlens occurs within a broadband of spatial frequency. A
superlens can be constructed from metamaterial with effective
negative index media\cite{Vesalgo68}\cite{Shelby01} consisting of
metallic resonators, and
dielectric\cite{Luo03}\cite{cubukcu03}\cite{Luo02}or
metallic\cite{Luo03ox} photonic crystals consisting of
periodically varying dielectric material. Also a superlens can be
built from a slab of natural material that can support surface
wave polaritons\cite{Fang05}\cite{Stockman05}\cite{Shalaev05} with
either negative permittivity or negative permeability.
Experimental studies have demonstrated superlens imaging both in
microwave regime\cite{Parimi03} using a two dimensional photonic
crystal and in optical regime using a silver
film\cite{Fang05}\cite{Lee05}\cite{Melville05}. Although the
resolution of the superlens is still limited by the internal
absorption losses of the materials due to inherent in strongly
dispersive material\cite{Garcia02}, imaging well beyond the
diffraction limit has been shown.

However, there is one drawback to the planar superlens. The
superlens images are still in near-field zone as demonstrated by
Podolskiy et al.\cite{Podolskiy05}, since the enhanced evanescent
waves remain evanescent after transmission and vanish very quickly
outside the superlens. Therefore, a challenge remains as how to
use the superlens effect to form an image in far field with a
resolution beyond the diffraction limit.

Another approach to recover the evanescent waves is to introduce
an antenna into the near field that interacts with the evanescent
waves and then radiates into the far field. In fact, this is the
fundamental principle of the near-field scanning optical
microscopy (NSOM) where optical nanoantennas such as plasmonic
nanoparticles or metallic tips are used. Considerable studies have
been devoted to interpret the far-field
signals\cite{Greffet97}\cite{carminati00}\cite{Greffet95}\cite{Porto00}\cite{Carney04}
depending of the NSOM configuration, and images with resolution
down to 10nm have been demonstrated possible. Nevertheless, NSOM
do not project a physical image as a conventional lens does, and
the slow scan speed prevents a dynamic imaging\cite{Hecht00},
often of practical importance.

In this letter, we show theoretically that a new device termed as
{\it far-field superlens} (FSL) can overcome the near-field
limitation of a superlens, that is, in other words, able to
project in far-field, an image of the evanescent component of the
near-field. Moreover, we demonstrate theoretically that the image
is unique. The image pattern is not a real space image, but rather
provides the field angular spectrum\cite{Goodman96}, i.e.
information on the object in spatial spectral domain.  The
far-field signal can be easily processed numerically in order to
obtain a real space image of the local field distribution above
the object with a resolution beyond the diffraction limit. As an
example, a realistic design of an optical FSL made of
metal/dielectric is proposed from exact numerical calculation.

\section{Imaging theory with a Far-field superlens made of arbitrary material}
Adding a periodic grating on a superlens positioned in the
near-field above an object may help to realize a lens-like imaging
with a resolution below the diffraction limit. However, the
imaging capability of a grating is not straightforward. Let us
first introduce some general transmission properties of a grating
without considering the superlens effect. We assume an object
radiating optical waves at a wavelength $\lambda_{0}$ below a
grating immersed into the same medium with a refractive index  .
For the sake of simplicity and without losing generality, we
consider a $2$ dimensional problem where the material properties
of both object and grating are function of $(x,z)$ and invariant
along the $y$ axis. The grating is periodic along the $x$ axis
with a periodicity $d$. Periodic gratings are known to be able to
couple out evanescent waves into propagating waves by a simple
diffraction process. This property can be understood by writing
the for instance the grating law:
\begin{equation}
k'=k+pG,
\end{equation}
where  $k'$ and $k$ are the transmitted and the incident
transverse wave number respectively; $p$ is the diffraction order;
and $G$ is the grating wave number of the grating. The transverse
wave number is the projection of the wavevector of a plane wave
along the $x$ axis. Transverse wave number of evanescent waves are
such that $|k|>nk_{0}$ where $k_{0}=2\pi/ \lambda_{0}$, while
transverse wave number of propagating waves satisfy to
$|k|<nk_{0}$. Incident evanescent waves with a large $k$  can be
lowered by the grating wave number using for instance the order
$p=-1$ of diffraction. Evanescent waves can be converted by this
way into propagating waves that reach the far-field if $G$ is
large enough. But incident propagating waves would be also
transmitted in far-field through the order $p=0$ without wave
number change. So that incident propagating and evanescent waves
transmitted through the order $0$ and $-1$ respectively will
overlap in far-field making it difficult to separate them for
imaging purposes. Indeed, waves transmitted in far-field for
instance with a transverse wave number $|k'|<nk_{0}$ are the
results of the overlap of incident waves transmitted through
several orders $p$ with transverse wave numbers satisfying:
\begin{equation}
k_{p}=k'-pG.
\end{equation}

Let us write the relationship between the field transmitted in
far-field and the incident field assuming TM polarized waves where
the magnetic field $H$ is oriented along the $y$ axis. The
$H$-field transmitted above the grating with  and its angular
spectrum\cite{Goodman96} are noted $H_{t}(x,z)$ and
$\widetilde{H}_{t}(k',z)$. Only plane waves with $|k'|<nk_{0}$
that can reach the far-field are considered. The near field
radiated by the object under the grating with $z_{0}<z<z_{1}$ is
noted $H_{obj}(x,z)$. The near-field can be decomposed into a
broadband angular spectrum of both propagating ($|k|<nk_{0}$) and
evanescent ($|k|>nk_{0}$) plane waves. Separated by a grating,
those two angular spectra are linked by a discrete linear
summation of waves scattered into all orders of diffraction:
\begin{equation}{\label{eq:transmission_}}
 \widetilde{H}_{t}(k',z_2)=\sum_{p=-\infty}^{+\infty}
 t_{p}(k_{p})\widetilde{H}_{obj}(k_{p},z_1),
\end{equation}
In Eq. (3), $t_{p}$ is the $p$-order field transfer function of
the grating from $z=z_{1}$ to $z=z_{2}$, defined as the ratio
between the field transmitted in the order $p$ of diffraction, and
the field of an incident plane wave. The transfer function is a
convenient tool commonly used in Fourier optics\cite{Goodman96} to
describe the transmission properties of optical system by plane
waves. Transfer functions can be either measured experimentally or
numerically by solving Maxwell's equations.

In general, the original near-field $\widetilde{H}_{obj}(k,z_1)$,
cannot be retrieved unambiguously from the far-field measurement
of the angular spectrum  $\widetilde{H}_{t}(k,z)$ using Eq. (2-3)
because of an overlap of several incident plane waves with
different $k_{p}$ scattered into with the same transverse wave
number $k'$ ( the same direction). In general, there is no
one-to-one relationship between the near-field angular radiated by
the object and the far-field angular spectrum transmitted by a
grating.

\begin{figure}
\includegraphics[width=6.5cm]{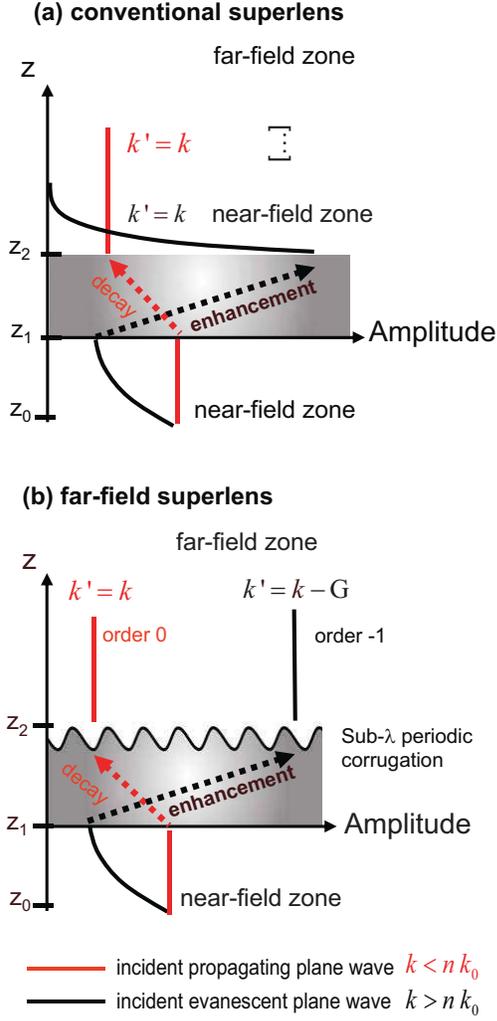}
\caption{\label{fig:1} Schematic picture of the transmission
properties of a conventional superlens versus a far-field
superlens. Through a conventional superlens (a), incident
evanescent waves are enhanced in transmission and vanish quickly
in the near-field zone. In contrast, a FSL (b) both enhances and
converts them into propagating waves by diffraction while blocking
incident propagating waves.}
\end{figure}

We demonstrate that this problem of the overlap of waves
transmitted through several order of diffraction can be overcome
by combining the superlens effect to the diffraction properties of
a grating. The imaging principles can be understood following a
very simple picture. Let us look first at transmission properties
of a planar superlens as show in Fig 1a. Transmitted amplitude of
incident evanescent waves (in black) are substantially enhanced
through the slab because of the superlens effect. Comparatively
incident propagating waves are poorly transmitted (in red).
However, after transmission enhanced evanescent waves remain
evanescent, limiting imaging with a superlens in the near-field
zone. In contrast, let us consider now transmission properties of
planar superlens corrugated with a subwavelength periodic
structure termed as {\it far-field superlens} (FSL). As shown in
Fig. 1b, a FSL not only enhances the incident evanescent field
because of the excitation of surface waves in the slab based
superlens, but also effectively convert these waves into
propagating waves by scattering through a negative diffraction
order of the grating following Eq 1. Overall, these incident
evanescent waves transmitted and converted into propagating waves
are projected in the far-field with large amplitude. In the others
hand compared to the transmission of incident evanescent waves,
incident propagating waves are expected to be very poorly
transmitted in far-field because of a lack of surface wave
excitation, This property may be written:
\begin{equation}\label{eq:fsl_condition}
|t_{0}(k-G)|<<|t_{-1}(k)|,
\end{equation}
with $|t_{0}(k-G)|<<|t_{-1}(k)|$ within the bandwidth of
evanescent waves for which the superlens effect occurs. Let us
note that a similar relation occurs for negative transverse wave
numbers if the grating has a -axis symmetry grating. If in
addition, the superlens is designed with a large transmission
within selective bandwidth $k\in [G;nk_{0}+G]$, then the
relationship between the far-field angular spectrum above the FSL
and the near-field angular spectrum below the superlens given by
Eq (2) and (3) reduces to:
\begin{equation}
\widetilde{H}_{t}(k',z_2)=t_{-1}(k)\widetilde{H}_{obj}(k,z) \text{
where } k=k'+G,
\end{equation}
for the positive half-space $0<k'<nk_{0}$; and
\begin{equation}
\widetilde{H}_{t}(k',z_1)=t_{+1}(k)\widetilde{H}_{obj}(k,z) \text{
where } k=k'-G,
\end{equation}
It follows from this result that any propagating wave transmitted
in far-field by a FSL has a unique origin. For a positive $k'$ for
instance, the origin is the incident evanescent wave that has been
transmitted through the diffraction order $-1$ with $k=k'+G$. This
property is true for any $k'$ so that there is a unique one to one
relationship between the near-field angular spectrum under the FSL
and the transmitted angular spectrum in far-field above the FSL.
This results means that when an object is placed in close
proximity of a FSL, a unique image of the near-field distribution
can be projected in far-field. Moreover, using Eq. (5) and (6) and
the rigorous diffraction theory\cite{Goodman96}, the near-field
angular spectrum radiated by the object can be retrieved
unambiguously from measurement of the far-field transmitted
angular spectrum $\widetilde{H}_{t}(k',z)$.

If both amplitude and phase of the angular spectrum can be
measured in far-field, then a real space image of the near-field
$\widetilde{H}_{obj}(k,z)$ above the object can be reconstructed
from $H_{obj}(x,z)$ using a simple inverse Fourier transform.
However, the measurement of the phase is a practical difficulty.
This difficulty appears also in diffraction optical
microscopy\cite{Lauer02} where both amplitude and phase of the
angular spectrum have to be measured. For this purpose, an
experimental set-up such as the one use by Lauer\cite{Lauer02}
based on interferometry may be a good approach. Alternatively, a
direct real space image might be obtained using the Fourier
transform transmission properties of lens\cite{Goodman96} and
other optical devices.

In principle, the maximum spatial frequency of the electromagnetic
field that a FSL can image in far-field is
$(n+\lambda_{0}/d)k_{0}$. Consequently, the best transverse
resolution $\Delta l$ that could be obtained on the image of the
local density of electromagnetic energy is:
\begin{equation}
\Delta l=\frac{\lambda_{0}}{2(n+\lambda_{0}/d)}.
\end{equation}
By comparison, the best resolution that could be achieved with a
diffraction limited microscope is $\lambda_{0}/2n$ assuming a
numerical aperture NA=n.

Using a FSL, we have demonstrated that the near-field angular
spectrum and subsequently the local near-field distribution can be
measured. However, the electromagnetic distribution of the field
above the object depends on how the object is exposed. For
instance, in normal incidence or with a grazing angle exposure by
a plane wave, the FSL would provide accordingly different images.
A model is needed if one wants to image an intrinsic property of
the object that does not depend on the exposure condition such as
the local polarizability or the local absorptivity.

\section{Case of a silver far-field superlens}
\begin{figure}
\includegraphics[width=8.5cm]{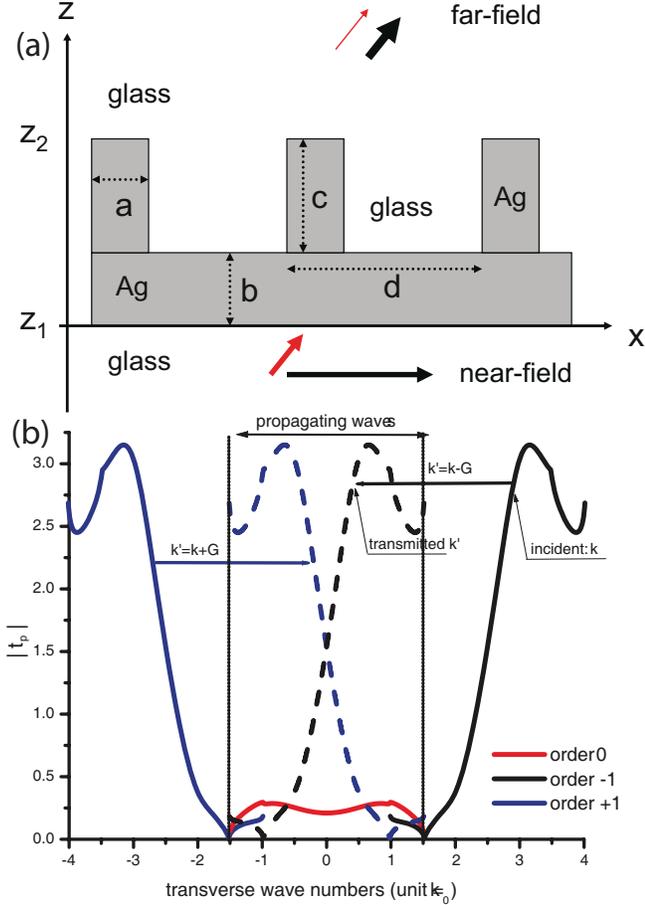}
\caption{\label{fig:2} (a) Design of an optical FSL working at
$\lambda_{0}=376nm$ with $a=45nm$ ; $b=35nm$ ; $c=55nm$ ;
$d=150nm.$ (b) Amplitude of transmission factor through order
$p=0$ (red) and order $p=-1$ (black) from near-field $z=z_{1}$ to
far-field ($z>>\lambda_{0}$) of the optical FSL shown in (a). This
FSL satisfies the requirement for imaging purpose: it provides a
strong transmission of evanescent waves and convert them into
propagating waves through the order -1 while the transmission of
propagating waves through the order 0 is comparatively small.}
\end{figure}

How to design such a FSL for which the transmission properties
satisfy to Eq (4) is a crucial question. One may start from the
design of a superlens slab that enhances strongly incident
evanescent waves within a large bandwidth. The enhancement can be
provided by the excitation of surface waves mode of the slab based
superlens. When a periodic corrugation is added on a superlens,
the surface modes supported by the slab become leaky modes. As a
result, as it was demonstrated by Smith et al\cite{Smith03} in
case of a superlens made of metamaterial with a negative
refractive index, corrugations at the interfaces of the superlens
lead to smaller values of the enhancement of evanescent waves by
the superlens. Despite this expected difficulty, we have
successfully designed an optical FSL made of silver/glass with the
proper transfer functions satisfying to Eq (4), a necessary
condition for imaging purpose. Details on the design of this FSL
are provided in Ref.\cite{DURANT06}. Feature sizes of this
nanostructure are shown in Fig 2a. This FSL has been designed to
work at $\lambda_{0}=376nm$ with TM polarized waves. We have
computed the transfer functions of this structure from $z=z_{1}$
to the top $z=z_{2} $ by solving numerically Maxwell's equations
using the Rigorous Coupled Wave Analysis (RCWA)
algorithm\cite{Moharam95_1}\cite{Moharam95_2} with experimental
permittivity data of glass $\epsilon=2.31$ and
silver\cite{Johnson72} $\epsilon=-3.16+0.2i$. The numerical
solution provided has been tested using the theorem of reciprocity
of electromagnetic waves and was applied for both propagating and
evanescent waves\cite{Carminati00_2}\cite{Carminati98}. The
results of order $0$ and $\pm 1$of the amplitude transfer
functions are plotted in Fig. 2b. With a periodicity $d=150nm$ and
the wavelength $\lambda_{0}=376nm$, incident transverse wave
numbers transmitted through the order $-1$ of the grating are
shifted by $-2.5k_{0}$.

Fig. 2b clearly shows that Eq (4) is satisfied with
$k\in[2.5;4]k_{0}$, demonstrating that using a superlens
periodically corrugated, a large bandwidth of evanescent waves can
be both enhanced and converted into propagating waves with large
amplitude, while incident propagating waves are poorly
transmitted. Consequently, this FSL could be used for imaging with
resolution well below the diffraction limit. Fig 2b shows
similarly that $|t_{0}(k+G)|<<|t_{+1}(k)|$ with
$k\in[-4;-2.5]k_{0}$ .

In a superlens made of silver, surface plasmon
polaritons\cite{Barnes03} (SPP) play a key role on the enhancement
of evanescent waves \cite{Smith05}\cite{Fang05}\cite{Lee05}. At a
metal/dielectric interface, SPP are surface waves resulting from
the coupling between p-polarized electromagnetic waves and the
induced collective excitation of free electrons in the metal. In a
silver film superlens, the wavelength and the thickness are chosen
so that SPP can be excited within a large bandwidth of transverse
wave numbers\cite{Fang05}\cite{Lee05}. How the optical FSL
presented in this letter has been designed in close connection to
SPP behavior, is detailed in Ref.\cite{DURANT06}.

Due to the position of the selective bandwidth of enhancement as
shown Fig. 2b, waves transmitted into order $-1$ and $+1$ can be
substantially overlapped with $k\in[-0.2;0.2]$. For this reason,
this small bandwidth has to be omitted from the measurement in
order to retrieve the near-field angular spectrum unambiguously.
Finally, it can be deduced using Eq (5) and (6) that the near
field angular spectrum $\widetilde{H}_{obj}(k,z)$ with
$k\in[-4;-2.7]\cup[2.7;4]k_{0}$ can be retrieved from the
measurement of the far-field angular spectrum
$\widetilde{H}_{t}(k',z)$ with
$k'\in[-1.5;-0.2]\cup[0.2;1.5]k_{0}$. Because this specific FSL
can resolve a transverse field modulation with a maximum spatial
frequency $4k_{0}$ , the transverse resolution on the image of the
local density of electromagnetic energy is $\lambda_{0}/8$. By
comparison, the resolution of diffraction limited microscope is
$\lambda_{0}/3$ with a numerical aperture $NA=1.5$.

\begin{figure}
\includegraphics[width=8.5cm]{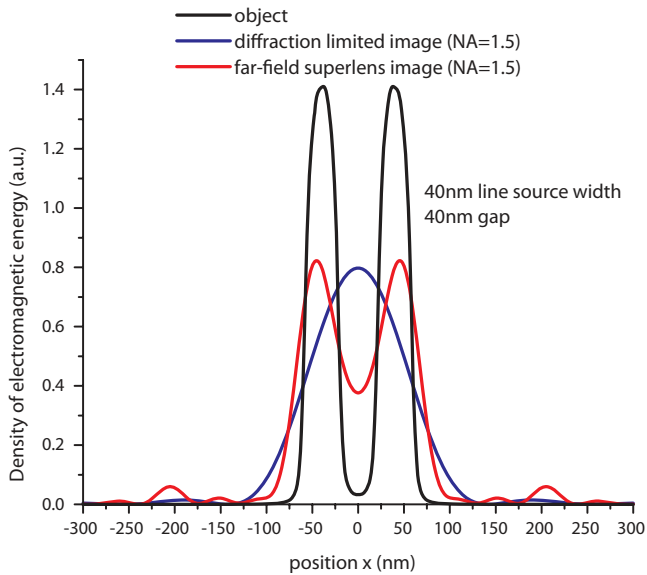}
\caption{\label{fig:3} Electromagnetic energy 5nm above the object
and corresponding images with and without FSL assuming in both
case a numerical aperture NA=1.5. Image of the density of
electromagnetic energy with the FSL is reconstructed using Eq 5
and 6 and from the rigorous computation of the transmitted angular
spectrum in far-field $z>>\lambda_{0}$. This result computed
rigorously directly demonstrates the optical imaging method with
resolution below the diffraction limit from Far-field data, using
a FSL made of silver/glass without any scanning}
\end{figure}

As an example, we provide the result of the image reconstruction
using a FSL, of an object constituted of two 40nm lines sources of
coherent TM waves separated by a deep subwavelength gap. A unity
value of the $H$ component is assumed on the two-line sources and
vanished everywhere else. The near field computed at 5nm above the
object using rigorous diffraction theory\cite{Goodman96}, is shown
as the black curve in Fig. 3. For comparison, the computed
diffraction limited image using conventional optical microscope
assuming a numerical aperture $NA=1.5$ is shown in blue in Fig 3.
The optical FSL described in Fig 2a is placed 20nm above the
object. The angular spectrum $\widetilde{H}_{t}(k',z)$ transmitted
by the FSL in far-field ($z>>\lambda_{0}$) is computed rigorously
using RCWA and Eq. (2-3). Values of $\widetilde{H}_{t}(k',z)$
provide a complete set of data simulating a measured signals in an
experiment. These known data are used subsequently for the image
reconstruction. Because the designed silver FSL processes a unique
one to one $k'\rightarrow k$ relation, it allows us to use this
"experimental" data to retrieve the angular spectrum of the
near-field 5nm above the object unambiguously only by using
Eq.(5-6) and the rigorous diffraction theory. By combining this
angular spectrum with its propagating component, we obtain the
near-field angular spectrum 5nm above the object with
$k\in[-4;-2.7]\cup[-1.5;1.5]\cup[2.7;4]k_{0}$. By applying a
simple inverse Fourier transform, we finally obtain successively a
reconstruction of the $H$-field distribution and the image of the
density of electromagnetic energy 5nm above the object. We have
successively obtained faithful images reconstruction from 120nm
down to 30nm gap. The case of 40nm gap is reported by the red
curve in Fig. 3 where the separation of the two lines source is
very clear. Let us note the formation of some
artefacts\cite{DURANT06} in the image may appear because of the
missing band in the near-field angular spectrum. The image in case
of 30nm gap (not shown) is the smallest gap between sources that
we have been able to demonstrate following the Rayleigh
criterion\cite{Born80}.

\section{Conclusion}
We have demonstrated theoretically how to overcome the limitation
of a conventional superlens for which only images in the
near-field can be obtained\cite{Fang05}\cite{Podolskiy05}. We have
shown that when the object is positioned close to a new device
termed as the far-field superlens (FSL), a unique image of
evanescent waves radiated by the object can be formed in
far-field. In contrast to conventional near-field scanning optical
microscope (NSOM), the FSL does not require scanning. In this
sense, a FSL is similar with conventional lenses imaging with
which a whole and unique image of an object can be recorded in a
single snap-shot. From the measurement of the far-field image
pattern and a simple inversion of the linear and scalar Eqs. (5)
and (6), the near-field electromagnetic distribution above the
object can be obtained with a resolution beyond the diffraction
limit. By combining the superlens effect and the diffraction modes
of a grating, the unique transmission properties of a FSL lies in
a broadband excitation of surface wave leaky modes used to convert
the incident near-field angular spectrum into a transmitted
far-field angular spectrum, related by a one to one relationship.
A realistic design of an optical FSL was given made of
silver/glass with such a transmission properties, owing to the
excitation of surface plasmon polariton (SPP) leaky modes. This
new imaging approach has the potential to reach similar or better
resolution than NSOM after more development. Such a far field
superlens could have great impact not only in nano-scale imaging
but also in nanolithography and photonic devices.

\begin{acknowledgments}
We are very grateful to Dr Q.-H. Wei for the critical reading of
the manuscript. This research was supported by NSF Nano-scale
Science and Engineering Center (DMI-0327077) and ONR/DARPA
Multidisciplinary University Research Initiative (MURI)
(Grant\#N00014-01-1-0803).
\end{acknowledgments}

\newpage 
\bibliography{FSL}

\end{document}